\documentclass[aps,12pt,notitlepage,nofootinbib]{revtex4-1}
\usepackage{graphicx}
\usepackage{cancel}
\usepackage{textcomp}
\usepackage{amsmath, amssymb, amsfonts, latexsym, epsfig}
\usepackage{mathrsfs}

\usepackage{bm}
\usepackage{times}
\usepackage{epsfig}
\usepackage{color}

\def\beq{\begin{equation}}
\def\eeq{\end{equation}}
\def\be{\begin{equation}}
\def\ee{\end{equation}}
\def\bea{\begin{eqnarray}}
\def\eea{\end{eqnarray}}
\def\to{\rightarrow}

\begin{document}
\title{Higgs Precision Measurements and Flavor Physics: \\ A Supersymmetric Example\\
\vspace{0.8cm}
\underline{\it Review for ``Chinese Science Bulletin'' and CEPC+SPPC Proposal}
}
\author{Kai Wang}
\author{Guohuai Zhu}
\affiliation{ Zhejiang Institute of Modern Physics and Department of Physics, Zhejiang University, Hangzhou, Zhejiang 310027, CHINA}

\begin{abstract}
Rare decays in flavor physics often suffer from Helicity suppress and Loop suppress. 
Helicity flip is a direct consequence of  chiral $U(3)$ symmetry breaking and electroweak symmetry breaking.
The identical feature is also shared by the mass generation of SM fermions. In this review, we use MSSM
as an example to illustrate an explicit connection between bottom Yukawa coupling and rare
decay process of $b\to s\gamma$. We take a symmetry approach to study the common symmetry breaking in supersymmetric correction to bottom quark mass generation and $b\to s\gamma$. We show that 
Large Peccei-Quinn symmetry breaking effect and $R$-symmetry breaking effect required by $b\to s\gamma$ inevitably lead to significant reduction of bottom Yukawa ${y}_{b}$. To compromise the reduction in $b\bar{b}$,
a new decay is also needed to keep the Higgs total width as the SM value. 
\end{abstract}

\maketitle

\section{Introduction: chiral and electroweak symmetry breaking}
A SM-like Higgs boson has been discovered at both ATLAS and CMS detectors at the CERN Large Hadron Collider first via the two cleanest channels, di-photon and  four-lepton, with reconstructed invariant mass of 125~GeV\cite{today}.  The di-lepton mode was also seen with mass range consistent with the four-lepton measurement \cite{ww}. Updates from the two collaborations \cite{Atlas-Spin,CMS-Spin} also prefer a CP-even spin-zero state $J^{PC}=0^{++}$. The over-5 $\sigma$ evidence of $gg\to \phi\to ZZ^{*}\to \ell^{+}_{i}\ell^{-}_{i}\ell^{+}_{j}\ell^{-}_{j}$ clearly indicates that the boson $\phi$ is responsible for electroweak symmetry breaking and should be identified as the Higgs boson. In addition, both collaborations have also reported the boson decaying into tau pairs, $\phi\to \tau^{+}\tau^{-}$ which is the first evidence at the LHC that the Higgs-like boson actually couples to SM fermions. However, the final confirmation of whether the Higgs boson is the standard model (SM) Higgs boson still requires precision measurement of the Higgs couplings. For instance, in the so-called ``decoupling limit'', many new physics models beyond the SM also predict a light Higgs boson with couplings only differ from the SM ones by 10\% or less. There is also example where
the other couplings of this Higgs boson except the bottom Yukawa are very similar to the SM Higgs couplings while the bottom Yukawa measurement itself still suffer from
large uncertainty.  Higgs physics has entered an era of precision measurement and various $e^{+}e^{-}$ colliders as Higgs factory have been proposed as 
one intensity frontier with controlled background to improve the measurement of Higgs couplings. 

On the other hand, the other type of intensity frontier as flavor factories have been 
playing important role in searching physics beyond the SM for many years.  In
this review, we try to illustrate the direct correlation between physics at two types of intensity frontier, the Higgs precision measurement and flavor physics. 

Fermion mass is a consequence of chiral symmetry breaking. The Lagrangian of the kinetic energy and gauge interactions
of the SM fermion fields is
\beq
 i\bar{Q}^{i}_{L}\cancel{D}Q^{i}_{L} + i\bar{u}^{i}_{R}\cancel{D}u^{i}_{R}+ i\bar{d}^{i}_{R}\cancel{D}d^{i}_{R}
+  i\bar{\ell}^{i}_{L}\cancel{D}\ell^{i}_{L}+i\bar{e}^{i}_{R}\cancel{D}e^{i}_{R}\label{lagrangian}
\eeq
where index $i$ stands for generation. In Eq.\ref{lagrangian}, all the fields carry unbroken gauge symmetry $U(1)_{EM}$ and some are fundamental representation
of $SU(3)_{C}$, Majorana masses are strictly forbidden for the above fields. 
Lagrangian in Eq.\ref{lagrangian} is also invariant under global unitary transformations
\beq
f^{i} \to U^{ij}_{f} f^{j}, f^{i}\in \{Q^{i}_{L}, u^{i}_{R},d^{i}_{R},\ell^{i}_{L},e^{i}_{R}\}
\eeq
which correspond to accidental chiral symmetries $U(3)_{Q}\times U(3)_{u}\times U(3)_{d}\times U(3)_{\ell}\times U(3)_{e}$ for three generations. 
Yukawa couplings $y_{u,d,e}$ of the SM fermions to the Higgs boson, 
\beq
-y^{ij}_{u} \bar{Q}^{i}_{L} u^{j}_{R}\bar{H}-y^{ij}_{d}\bar{Q}^{i}_{L} d^{j}_{R}H-y^{ij}_{e}\bar{\ell}^{i}_{L}e^{j}_{R}H+h.c.,  \bar{H}=\epsilon H^{*},
\eeq
explicitly break the above
chiral symmetries and SM fermion masses arise after $H$ developing the vacuum expectation value. Therefore, SM fermion mass generation is a consequence of both chiral symmetry breaking and electroweak symmetry breaking.

Rare decays in flavor physics often suffer from Helicity suppress and Loop suppress.
For instance, pseduo-scalar leptonic decay $\pi^{-}\to e^{-}\bar{\nu}_{e}$ is suppressed by the electron mass. The SM contribution to $B_{s}\to\mu^{+}\mu^{-}$ is suppressed by the muon mass insertion. Dipole operator 
\beq
\bar{b}\sigma_{\mu\nu} s F^{\mu\nu}
\eeq
which correspond to $b\to s\gamma$. Existence of on-shell spin-one photon implies that the helicity in involving quark states must be flipped in $b\to s\gamma$ process. 
Helicity flip also breaks chiral symmetries and electroweak gauge symmetry. 
In SM, the helicity flip here corresponds to a mass insertion of $m_{b}$.
Therefore, there may exist a direct correlation between $b$-quark mass generation
and $b\to s\gamma$. In this review, we use MSSM as 
an example to illustrate the feature as how contribution to $b\to s\gamma$ may modify
the $b$ Yukawa coupling. 

\section{Type-II 2HDM and Peccei-Quinn symmetry}
We are interested in the deviation in bottom Yukawa coupling as
\beq
m_{b}=y_{b} v_{d} + \Delta m_{b}~.
\eeq
This is a typical feature Type-II Two-Higgs-Doublet-Model(2HDM) where quark mass
generation arises from two different electroweak symmetry breaking sources. 

Minimal Supersymmetric Standard Model (MSSM) is a natural Type-II 2HDM. The superpotential
being holomorphic so the $\bar{H}=\epsilon H^{*}$ is forbidden in superpotential.  The anomaly cancellation conditions for $[SU(2)_{L}]^{2}U(1)_{Y}$
and Witten anomaly also require the introduction of second Higgsino, the Fermonic partner of the Higgs boson, so that the Higgsino contributions to anomalies vanish.
MSSM superpotential is 
\beq
{\cal W}=y_{u}Qu^{c}H_{u}+y_{d}Qd^{c}H_{d}+y_{e}\ell e^{c} H_{d} + \mu H_{u} H_{d}
\eeq
The $\mu$ is a dimensional parameter which is constrained. $\mu$ cannot be zero to
avoid massless Higgsino and $\mu$ cannot be too large either so that the Higgs boson do not decouple. Suppose $\mu$ arise from a dynamical field $S$ which is SM singlet
\beq
{\cal W}\ni S H_{u} H_{d}~.
\label{hhs}
\eeq
In order to forbid the bare $\mu$-term in the superpotential, we assume there exists
a non-$R$ $U(1)_{X}$ symmetry under which $S$ transforms non-trivially $s\neq 0$. 
\bea
Qu^{c}H_{u}&:& q+ u+h_{u}=0\nonumber\\
Qd^{c}H_{d}&:& q+d+h_{d}=0\nonumber\\
SH_{u}H_{d}&:& s+h_{u}+h_{d}=0
\eea

\begin{figure}[ht]
\begin{center}
\includegraphics[scale=1,width=6cm]{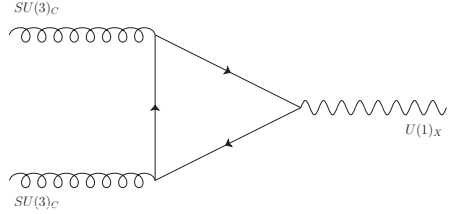}
\end{center}
\caption{Mixed QCD anomaly $A_{[SU(3)_{C}]^{2}U(1)_{X}}$.}
\label{anomaly}
\end{figure}

If one compute the mixed QCD anomaly with $U(1)_{X}$ as in Fig.\ref{anomaly},
one can obtain the anomaly coefficient 
\beq
A_{[SU(3)_{C}]^{2}U(1)_{X}} ={N_{f}\over 2} (2 q+ u+d) = -{N_{f}\over 2}(h_{u}+h_{d})={N_{f}\over 2} s\neq 0
\eeq
Therefore, a non-zero $s$ charge results in non-vanishing of mixed QCD-$U(1)_{X}$ anomaly $A_{[SU(3)_{C}]^{2}U(1)_{X}}$. The $U(1)_{X}$ can then be identified as
Peccei-Quinn (PQ) symmetry which is a global $U(1)$ symmetry with mixed QCD anomaly. The $\mu$-term which corresponds to the Higgsino mass term explicitly breaks the PQ symmetry. 

In a full MSSM, PQ symmetry breaking not only appears as Higgsino mixing $\tilde{H}_{u}\tilde{H}_{d}$ but also appear in scalar potential as  
\beq
V\ni \mid F_{H_{d}}\mid^{2}= y_{d}\mu^{*} H^{*}_{u}\tilde{Q} \tilde{d} +   y_{e}\mu^{*} H^{*}_{u}\tilde{\ell} \tilde{e}
\eeq
where 
\beq
F_{H_{d}}= {\partial W\over \partial H_{d}}= y_{d} Q d^{c} +y_{e} \ell e^{c} +\mu H_{u}~.
\eeq

When the global $U(1)$ PQ symmetry is broken 
by anomaly, a pseudo-Goldstone boson is generated with its
mass generated by non-perturbative QCD effect. 
The term in Eq.\ref{hhs} would lead to Weinberg-Wilczek axion which is excluded by $K\to \pi a$ constraint. Therefore, one can introduce an extra $S^{3}$ term to explicitly break the $U(1)_{\rm PQ}$ into $Z_{3}$  
known as NMSSM approach or exotic quarks
to cancel the above anomaly known as gauged $U(1)^{\prime}$ approach.
Coincidently, QCD invisible axion requires the PQ symmetry breaking scale  is the same as the intermediate scale in gravity mediation supersymmetric theory, $M^{2}_{\rm PQ}/M_{\rm Pl}\sim M_{\rm EW}$. The $U(1)_{X}$ can actually provide a simultaneous solution to strong CP problem and the $\mu$-term problem as Kim-Nilles mechanism based on supersymmetric DFSZ $H_{u}H_{d}S^{2}/M_{\rm Pl}$ axion model \cite{kimnilles}. 

\section{$R$-symmetry and Supersymmetric corrections to mass generation}

In analogy to the chiral symmetry breaking that is associated with fermion mass generation, the Majorana gaugino mass in supersymmetric theory is associated with  $R$-symmetry breaking.  A global $U(1)$ $R$-transformation is defined as a rotation over the anti-commuting coordinates (Grassmann variables)
$\theta$ and $\bar{\theta}$
\beq
R: \theta\to e^{i\alpha} \theta, \bar{\theta}\to e^{-i\alpha}\bar{\theta}
\eeq
Gauge vector superfields are real so they are neutral under $R$-transformation.
The gaugino component is then of $R$-charge 1 as
\beq
R: \lambda \to e^{i\alpha}\lambda
\eeq
and gaugino mass term ${1\over 2}M_{\lambda}\lambda\lambda$ always
break the $U(1)$ $R$-symmetry.

One can categorize the soft-supersymmetry breaking Lagrangian based on
the PQ and $R$ symmetries in 
\bea
{1\over 2}M_{\lambda}\lambda\lambda + \tilde{A}_{u}\tilde{Q}\tilde{u^{c}}H_{u}+...&:& \cancel{R}\nonumber \\
 B\mu H_{u} H_{d} &:& \cancel{\rm PQ}, \cancel{R} \nonumber \\
M^{2}_{\tilde{f}}\tilde{f}^{*}\tilde{f} &:&~.
\eea
The gaugino mass and $\tilde{A}$-terms break $R$-symmetry. $B\mu$-term
breaks both $R$ and PQ symmetries. The scalar mass term $\tilde{f}^{*}\tilde{f}$
is trivial under any unitary transformation. 
 
The two global $U(1)$ assignments in MSSM are not uniquely defined.  
In Table \ref{ccc}, we list one sets of assignment of PQ and $R$ charges 
consistent of $SU(5)$. 
\begin{table}[h]
\begin{center}
\begin{tabular*}{1.0\textwidth}{@{\extracolsep{\fill}} c || c c c   c c c c | c}
\hline
Field & $Q$ & $u^{c}$ & $e^{c}$ & $d^{c}$ & $\ell$ & $H_{u}$ & $H_{d}$ & $\theta$ \\
\hline\hline
$R$-charge &  ${1\over 5}$ & ${1\over 5}$ & ${1\over 5}$ & $7\over 5$ & ${7\over 5}$ & $8\over 5$ & $2\over 5$ & 1 \\
PQ   & 0 & 0 & 0 & -1 & -1 & 0 & 1 & 0 \\
\hline
\end{tabular*}
\label{ccc}
\caption{Charge assignment under $R$-symmetry and Peccei-Quinn symmetry.}
\end{center}
\end{table}

In MSSM, $b$-quark mass arises at the tree level from $y_{d}Qd^{c}H_{d}$.
The supersymmetric correction to $m_{b}$ is effectively   
\beq
Q u^{c} \bar{H}_{u}
\eeq
which is known to be Lorentz invariant and gauge invariant in SM. 
Using charge assignments in Table \ref{ccc}, one can substitute them into calculation of effective coupling $Q d^{c} \bar{H}_{u}$ as
\bea
R[Q d^{c} \bar{H}_{u}] : & {1\over 5}+{7\over 5} - {8\over 5} =0 \\
{\rm PQ}[Q d^{c} \bar{H}_{u}]: &  0+ (-1) +0 =-1~.
\label{pq}
\eea
Taking two Fermonic component,  the $R$-invariant condition 
is of $R$-charge 2 while the above term is 0 so it breaks $R$-symmetry
as well as  the PQ symmetry. A trivial realization is that the correction
can come from the two Higgs mixing term which is known as $B\mu$-term.
$B\mu$-term explicitly breaks the PQ and $R$ symmetries as discussed 
previously.
We can conclude that the supersymmetric correction to SM
fermion masses must break PQ and $R$ symmetries in addition
to the chiral symmetry and electroweak symmetry. 
Chiral symmetry breaking is quantized by the Yukawa coupling.
In the case of $m_{b}$, the correction is also proportional to $H_{u}$ {\it vev} $v_{u}$ which is typically dominated the electroweak symmetry breaking
since it is the dominant contribution to top quark mass.  
Besides $y_{b}$ and $v_{u}$, the size of correction $\Delta m_{b}$ is then proportional to scales that break the PQ and $R$ symmetries, including $B\mu$-term, product of $\mu$-term and gaugino masses or $A$-terms.

An inapparent electroweak symmetry breaking source is the Wino-Higgsino
mixing as in the neutralino mass matrix
\beq
N=
\left(
\begin{array}{cccc}
 M_{1} & 0   & -g_{1}v_{d}/\sqrt{2}   & g_{1}v_{u}/\sqrt{2} \\
 0 & M_{2}  & g_{2}v_{d}/\sqrt{2}   & -g_{2}v_{u}/\sqrt{2} \\
 -g_{1}v_{d}/\sqrt{2}  & g_{2}v_{d}/\sqrt{2}   & 0  & -\mu \\
 g_{1}v_{u}/\sqrt{2}  & -g_{2}v_{u}/\sqrt{2}   &    -\mu & 0
\end{array}
\right)~.
\eeq

\section{Non-decoupling MSSM an example}

To illustrate the feature, we take a non-decoupling limit \cite{cp,ours} where supersymmetric correction to $b\to s\gamma$ is maximized to cancel
the contribution from light charged Higgs.
In this limit, the Higgs boson of 125~GeV is identified as
the heavy neutral Higgs $H$ by taking $M_{A}$ around $m_{Z}$ scale 
 and the charged Higgs $H^{\pm}$ is also around ${\cal O}(100~\text{GeV})$ as the tree level contribution to charged Higgs mass
 $M^{2}_{H^{\pm}}= M^{2}_{A}+m^{2}_{W}$. Such a light charged Higgs
may significantly enhance the flavor violation $b\to s\gamma$. The 
2HDM constraint on $b\to s\gamma$ has pushed the charged Higgs
mass to be over 300~GeV. Significant cancellation to the light charged 
Higgs of ${\cal O}(100~\text{GeV})$  is then needed from supersymmetric
particles. As we argued, large supersymmetric correction is a consequence
of large PQ symmetry breaking and large $R$ symmetry breaking
so qualitatively it is easy to see where the allowed parameter region lies.  
At the same time, the inevitable supersymmetric correction  significantly modifies the bottom Yukawa coupling. Large reduction in $H\to b\bar{b}$ then results in large reduction of Higgs total width.
Since all other Higgs decay channels are at similar level as the SM predicts, new decay channel is then needed to compromise the reduction
of Higgs total width. This feature has been discussed in details by our
previous work \cite{ours} and we give a brief summary in this section.

\begin{figure}[h]
\includegraphics[scale=1,width=7cm]{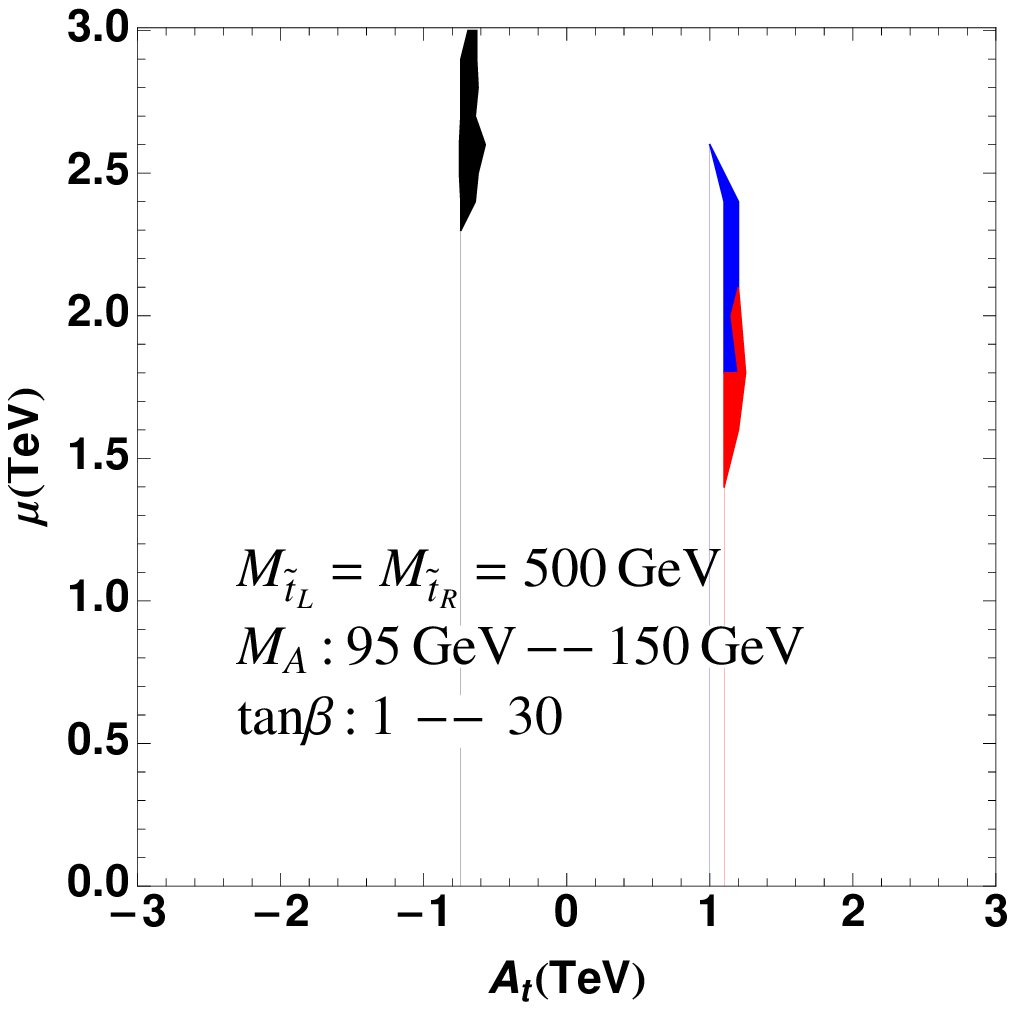}
\includegraphics[scale=1,width=7cm]{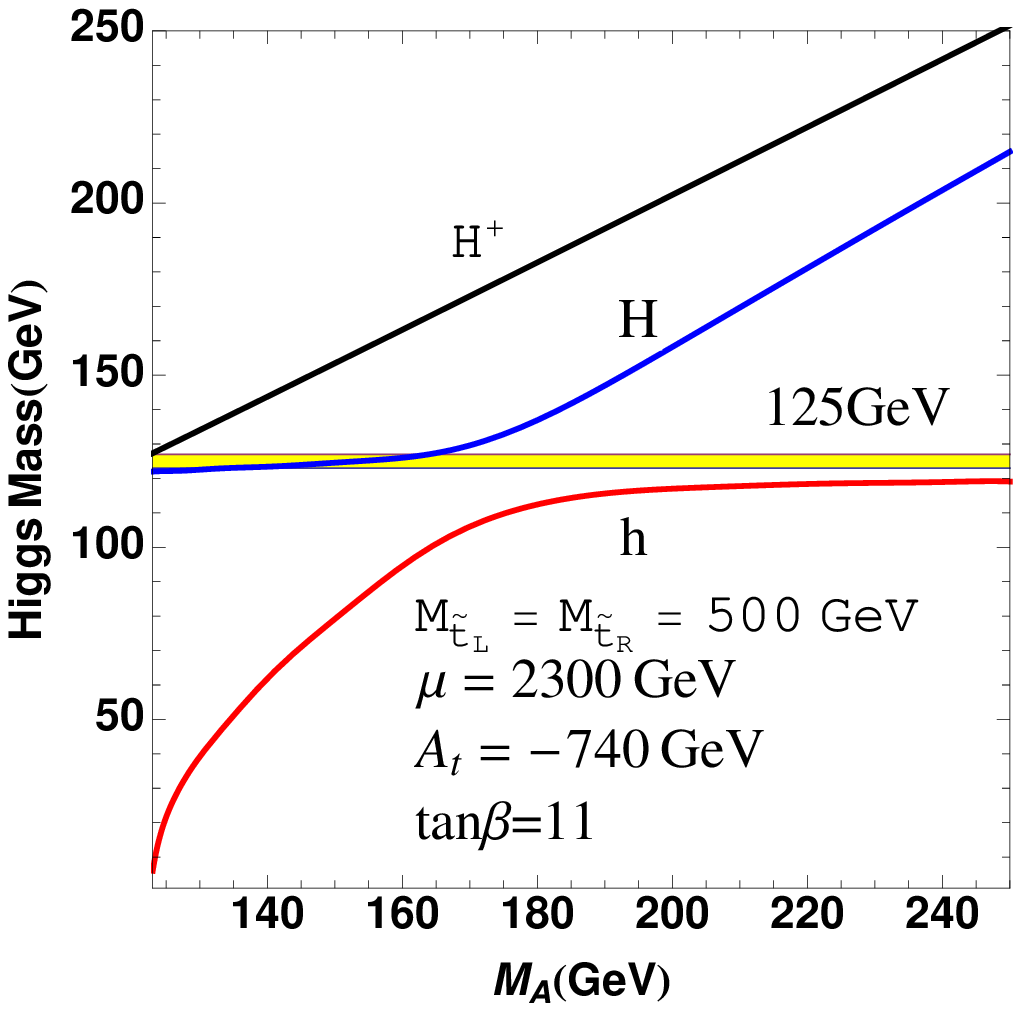}
\caption{(a) Scan Results in [$A_t$, $\mu$] plane. The heavy (light) stop scenario with $M_{\tilde{Q}_{3}}=M_{\tilde{t}}=1~(0.5)$~TeV is shown in the left (right) plot. The red region pass the direct search bounds from HiggsBounds with
a heavy CP-even Higgs $M_H=125 \pm 2$ GeV and an enhanced diphoton rate $1<R_{\gamma \gamma}<2$. The blue region pass in addition the constraint of BR$(B\to X_s \gamma)$, while the black region pass all the constraints, including further the restriction of BR$(B_s \to \mu^+ \mu^-)$. (b) $M_{h,H,H^{\pm}}$ vary with respect to $M_{A}$ for $M_{\tilde{t}}=500$~GeV, $A_{t}=-740$~GeV, $\tan\beta=11$, $\mu=2300$~GeV. }
\label{scan}
\end{figure}

In \cite{ours}, all the numerical analysis are performed with {\it FeynHiggs 2.9.2}~\cite{feynhiggs} with {\it HiggsBounds 3.8.0}~\cite{higgsbounds} and {\it SUSY\_Flavor 2.01}~\cite{Crivellin:2012jv}. We implement the requirements as
\begin{itemize}
\item $M_{H}: 125\pm 2$~GeV;
\item $R_{\gamma\gamma}=\sigma^{\gamma\gamma}_{\rm obs}/\sigma^{\gamma\gamma}_{\rm SM}: 1\sim 2$;
\item Combined direct search bounds from HiggsBound3.8.0;
\item BR$(B\to X_{s}\gamma)<5.5\times 10^{-4}$;
\item BR$(B_{s}\to \mu^{+}\mu^{-})<6\times 10^{-9}$~.
\end{itemize}
We also calculate the constraint
on $B^{+}\to \tau^{+}\nu$ and find the destructive interference between
the SM $W$ and the charged Higgs make the MSSM prediction about $20\% \sim 30\% $ smaller than the SM result of   $(0.95\pm0.27)\times 10^{-4}$. While the experimental world average is $(1.65\pm0.34)\times 10^{-4}$ before 2012 \cite{HFAG}, Belle updated their measurement at ICHEP2012 with much smaller value $0.72^{+0.29}_{-0.27} \times 10^{-4}$ for hadronic tag of $\tau$ \cite{Nakao}. So in the non-decoupling limit, a light charged Higgs with $\tan\beta \sim 10$ is well consistent with the new Belle measurement. In addition, the charged Higgs contribution to
$B\to D^{(*)}\tau\nu_{\tau}$ decays are not very significant in the interesting region of $M_{H^+}$ and $\tan\beta$.
In {\it FeynHiggs}, Higgs boson masses are calculated to full two-loop. To illustrate the qualitative feature here, we use the leading one-loop expression with only contributions of top Yukawa couplings.
Radiative corrections to the Higgs boson mass matrix elements and 
Higgs decay are \cite{Carena:1995bx,physicsreport,higherorder}. 
Figure \ref{scan} give the allowed parameter region for the fixed 
choice of top squark mass as 500~GeV to enhance the correction.
The points in blue region pass in addition the constraint of BR$(B\to X_s \gamma)$, while the points in black region pass all the constraints, including further the restriction of BR$(B_s \to \mu^+ \mu^-)$.  
It is clear that the survival region corresponds to large PQ symmetry and $R$ symmetry breaking where $\mu\sim 2--3$~TeV and $A_{t}\sim -750$~GeV. Figure \ref{scan}-b plots
the Higgs masses respect to $M_{A}$ for one set of benchmark points 
in the allowed parameter region. 

\begin{figure}[h]
\includegraphics[scale=1,width=7cm]{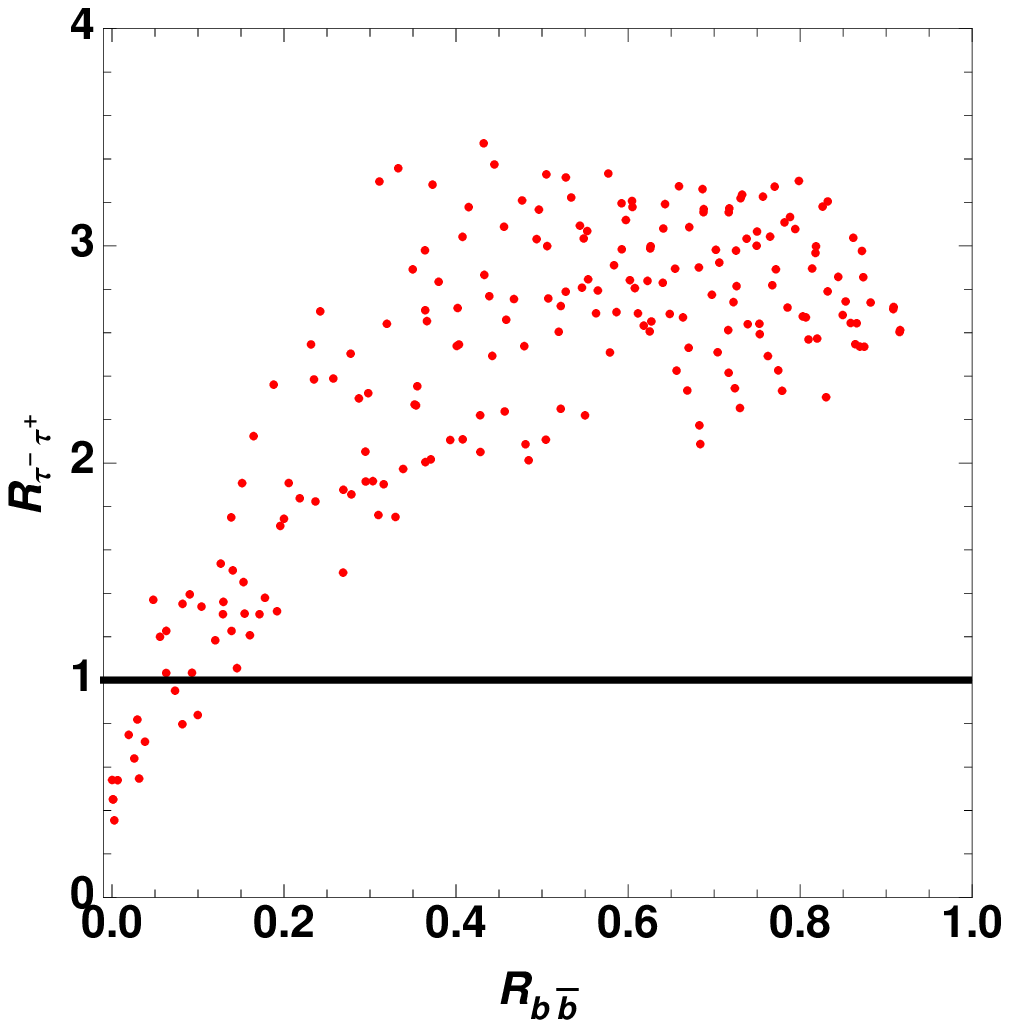}
\includegraphics[scale=1,width=8cm]{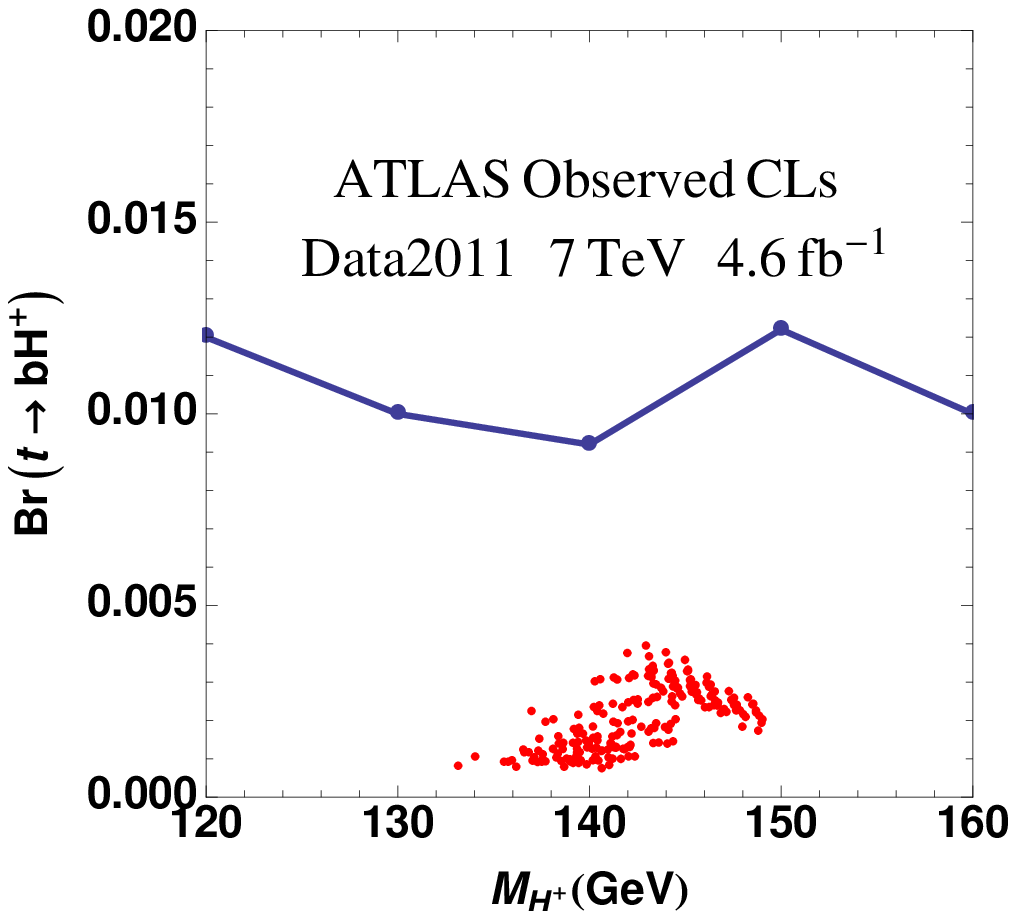}
\caption{(a) BR$(H\to\tau^{+}\tau^{-})$ in correlation with BR$(H\to b\bar{b})$ normalized by the SM values respectively. (b) BR$(t\to b H^{+})$ vs $M_{H^{\pm}}$ by assuming BR$(H^{+}\to \tau^{+}\nu_{\tau})=100\%$. Red dots are parameter points that pass all our selection and constraints.}
\label{tautau}
\end{figure}

When the supersymmetric correction in flavor physics processes is significant, bottom Yukawa also
receives significant corrections while at the same time $\tau$ Yukawa
is not modified significantly.  
Figure \ref{tautau} shows the correlation between BR$(H\to\tau^{+}\tau^{-})$ normalized by its SM value and BR$(H\to b\bar{b})$ normalized by the corresponding SM value as well as the corresponding BR$(t\to b H^{+})$ with respect to $M_{H^{\pm}}$ by assuming BR$(H^{+}\to \tau^{+}\nu_{\tau})=100\%$ 
for the survival points. 
In Fig.\ref{tautau}-a, the region where $R_{\tau^{+}\tau^{-}}\simeq 1$,
$H\to b\bar{b}$ partial width is much less than its SM value but 
a new decay of $H\to hh$ is opened and compromises the reduction
of $b\bar{b}$ partial width to make the Higgs total width remaining
unchanged.  
Fig.\ref{tautau}-b clearly shows that
all the parameter points that pass our selections are below the search of light charged Higgs boson via top decay $t\to b H^{+}$
with $H^{+}\to \tau^{+}\nu_{\tau}$.

\section{Conclusions}
Rare decays in flavor physics often suffer from Helicity suppress and Loop suppress. 
Helicity flip is a direct consequence of  chiral $U(3)$ symmetry breaking and electroweak symmetry breaking.
The identical feature is also shared by the mass generation of SM fermions
so one would expect that there exists a general correlation between
the helicity flip flavor physics process and SM fermion mass generation.
To illustrate the feature, we take a non-decoupling limit \cite{cp,ours} where supersymmetric correction to $b\to s\gamma$ is maximized to cancel
the contribution from light charged Higgs.
In this limit, the Higgs boson of 125~GeV is identified as
the heavy neutral Higgs $H$ by taking $M_{A}$ around $m_{Z}$ scale 
 and the charged Higgs $H^{\pm}$ is also around ${\cal O}(100~\text{GeV})$ as the tree level contribution to charged Higgs mass
 $M^{2}_{H^{\pm}}= M^{2}_{A}+m^{2}_{W}$. Such a light charged Higgs
may significantly enhance the flavor violation $b\to s\gamma$. The 
2HDM constraint on $b\to s\gamma$ has pushed the charged Higgs
mass to be over 300~GeV. Significant cancellation to the light charged 
Higgs of ${\cal O}(100~\text{GeV})$  is then needed from supersymmetric
particles. 
At the same time, the inevitable supersymmetric correction  significantly modifies the bottom Yukawa coupling. Large reduction in $H\to b\bar{b}$ then results in large reduction of Higgs total width.
Since all other Higgs decay channels are at similar level as the SM predicts, new decay channel $H\to hh$ is then needed to compromise the reduction
of Higgs total width. 

A more general approach to study correlation between Yukawa couplings and helicity flipped flavor violation operators is to appear in \cite{toappear}.
\section*{Acknowledgement}
KW is supported in part, by the Zhejiang University Fundamental Research Funds for the Central Universities (2011QNA3017) and the National Science Foundation of China (11245002,11275168). GZ is supported by the National Science Foundation of China (11075139,  and 11135006) and Program for New Century Excellent Talents in University.

\end{document}